\documentclass[prb,aps,twocolumn,amstex,preprintnumbers,amsmath,amssymb,array,tabularx]{revtex4}
\usepackage{graphicx}
\usepackage{dcolumn}
\usepackage{bm}
\usepackage{lscape} 
\usepackage{amsmath}
\usepackage{amsxtra}
\usepackage{tabularx}

\DeclareGraphicsExtensions{.jpg,.pdf,.eps,.gif}
\newcommand{\sml}[1]{ \scriptscriptstyle #1 }

\begin{document}

\title{Enhanced Half-Metallicity in Edge-Oxidized Zigzag Graphene Nanoribbons}

\author{Oded Hod, Ver\'onica Barone, Juan E. Peralta, and Gustavo E. Scuseria}

\affiliation{Department of Chemistry, Rice University, Houston, Texas 77005-1892}

\date{\today}

\begin{abstract}
We present a novel comprehensive {\it first-principles} theoretical
study of the electronic properties and relative stabilities of
edge-oxidized zigzag graphene nanoribbons.  The oxidation schemes
considered include hydroxyl, carboxyl, ether, and ketone groups.
Using screened exchange density functional theory, we show that these
oxidized ribbons are more stable than hydrogen-terminated nanoribbons
except for the case of the etheric groups.  The stable oxidized
configurations maintain a spin-polarized ground state with
antiferromagnetic ordering localized at the edges, similar to the
fully hydrogenated counterparts.  More important, edge oxidation is
found to lower the onset electric field required to induce
half-metallic behavior and extend the overall field range at which the
systems remain half-metallic.  Once the half-metallic state is
reached, further increase of the external electric field intensity
produces a rapid decrease in the spin magnetization up to a point
where the magnetization is quenched completely.  Finally, we find that
oxygen containing edge groups have a minor effect on the energy
difference between the antiferromagnetic ground state and the
above-lying ferromagnetic state.

\end{abstract}

\maketitle
 


Low-dimensional carbon structures such as fullerenes~\cite{Kroto1985}
and carbon nanotubes~\cite{Ijima1993} (CNTs) are promising candidates
for building blocks of future nanoelectronic and nanomechanical
devices.~\cite{dress-book, de-Heer2002} Made of a unique hexagonal
carbon lattice confined to a quasi one-dimensional (1D) tubular
structure, CNTs may be either semiconducting or metallic, depending on
their diameter and chirality.\cite{saito_book} Combined with their
ballistic~\cite{Tans1997, White1998, Frank1998, Bachtold2000,
White2001, Liang2001} electronic transport characteristics, this opens
exciting possibilities for the design of novel electronic components
and interconnects with nanometer scale dimensions.

Recently, a new type of graphene-based material was experimentally
realized.~\cite{Geim2004} Shaped as narrow stripes cut out of a single
(or a few) layer(s) of graphite, these elongated materials were named
graphene nanoribbons (GNR).  Since they share the same hexagonal
carbon lattice structure, GNRs and CNTs exhibit many similarities with
respect to their electronic properties.  However, the planar geometry
of the ribbons allows for the application of standard lithographic
techniques for the flexible design of a variety of experimental
devices.~\cite{Geim2004, Zhang2005, deHeer2006, Novoselov2007,
Han2007} The utilization of these established fabrication methods
suggest the possibility of a controllable and reproducible fabrication
of carbon based electronic components at the nanometer scale.

Despite the aforementioned similarities, there is a distinct
difference between CNTs and GNRs.  Unlike CNTs, GNRs present long and
reactive edges prone to localization of electronic edge states and
covalent attachment of chemical groups~\cite{Ramprasad1999} that can
significantly influence their electronic properties.
\input{epsf}
\begin{figure}[h]
\begin{center}
\begin{tabular}{c c c c c c}
\multicolumn{6}{l}{Periodic Direction} \\
\multicolumn{6}{l}{{\Huge $\longrightarrow$}} \\
\epsfxsize=1.18cm \epsffile{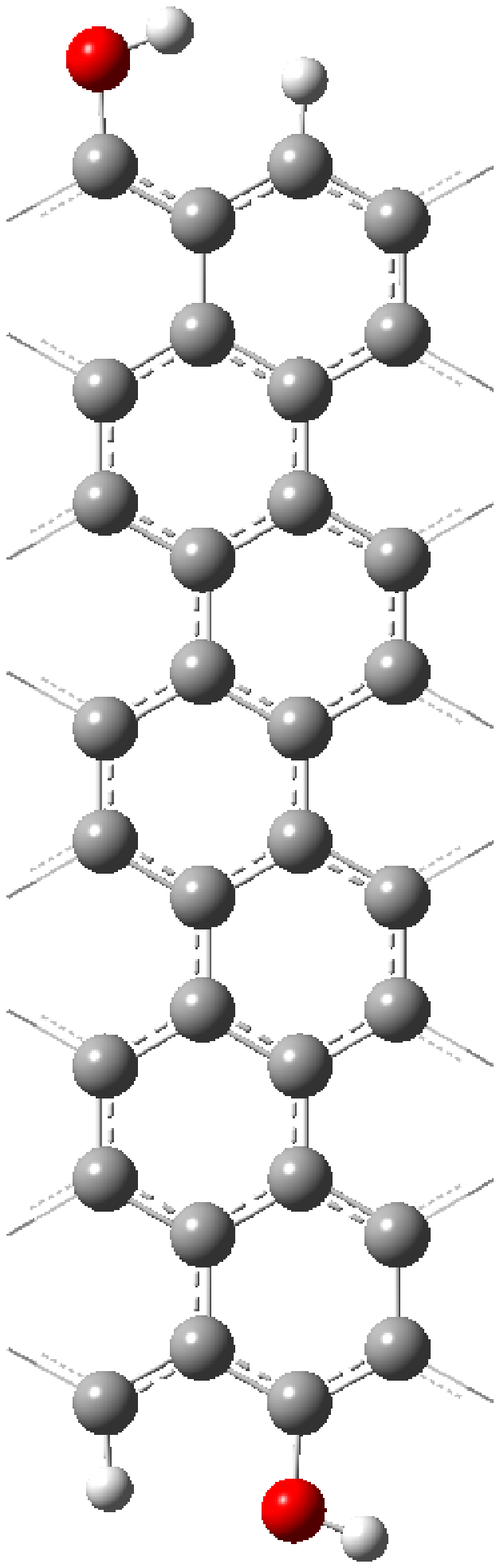}     &
\epsfxsize=1.20cm \epsffile{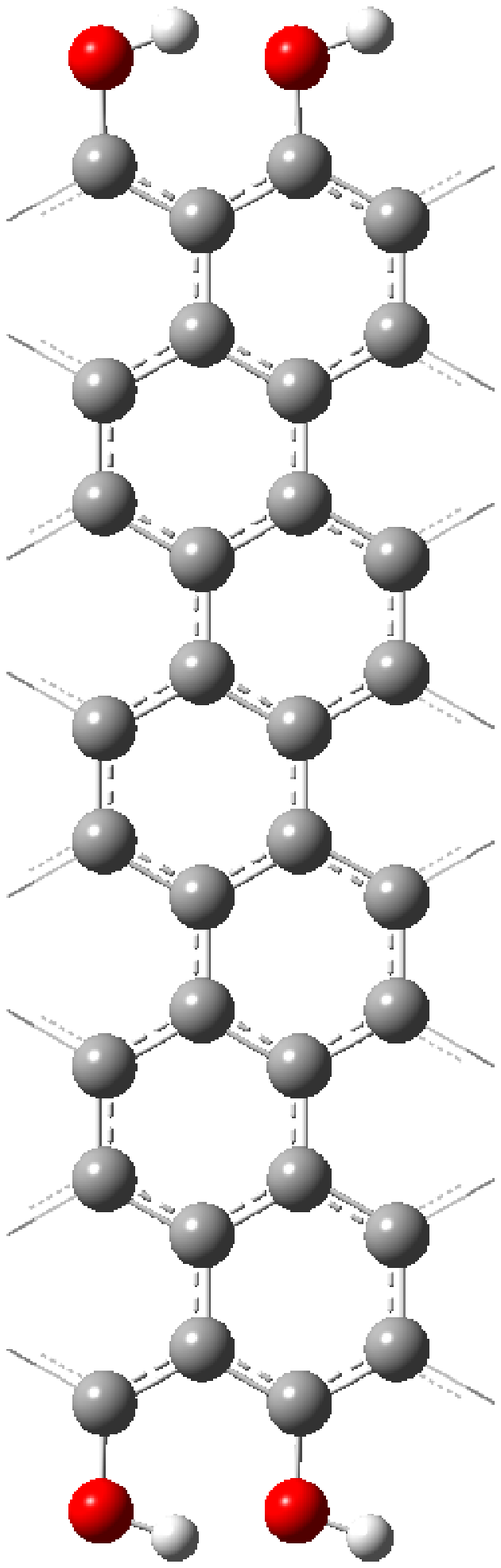}     &
\epsfxsize=1.10cm \epsffile{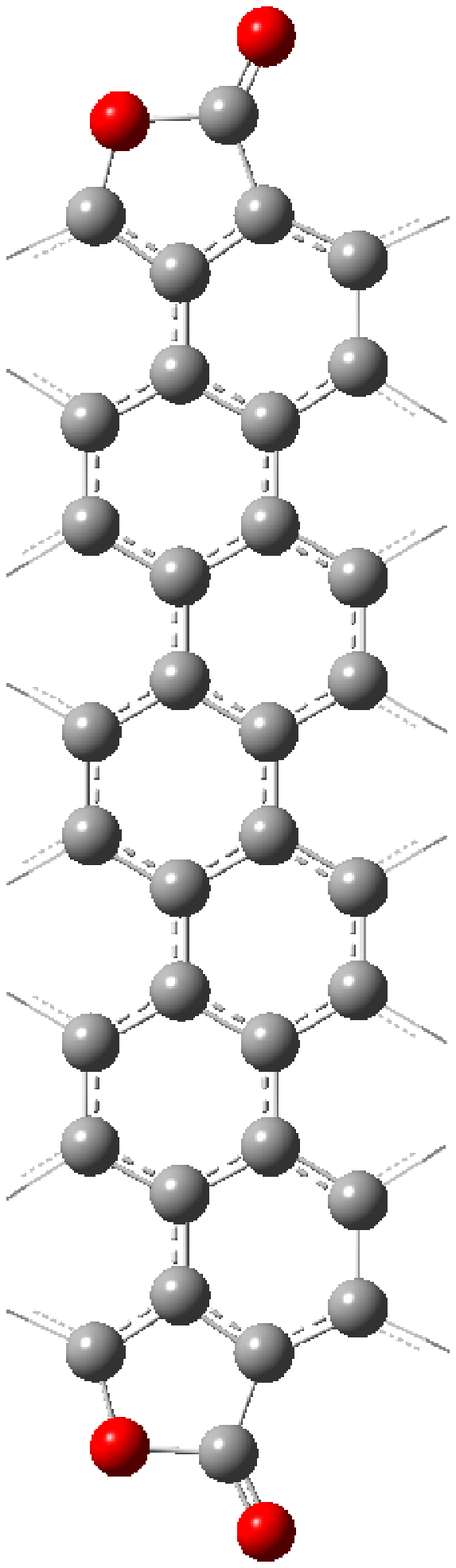}     &
\epsfxsize=1.20cm \epsffile{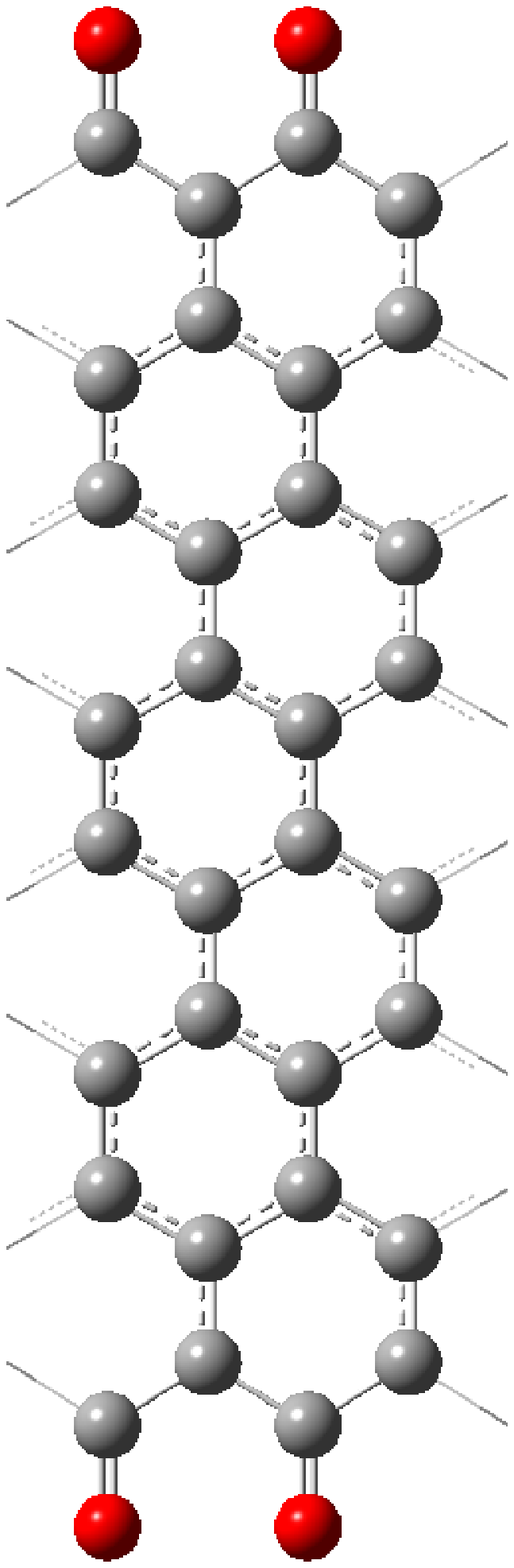}      &
\epsfxsize=1.24cm \epsffile{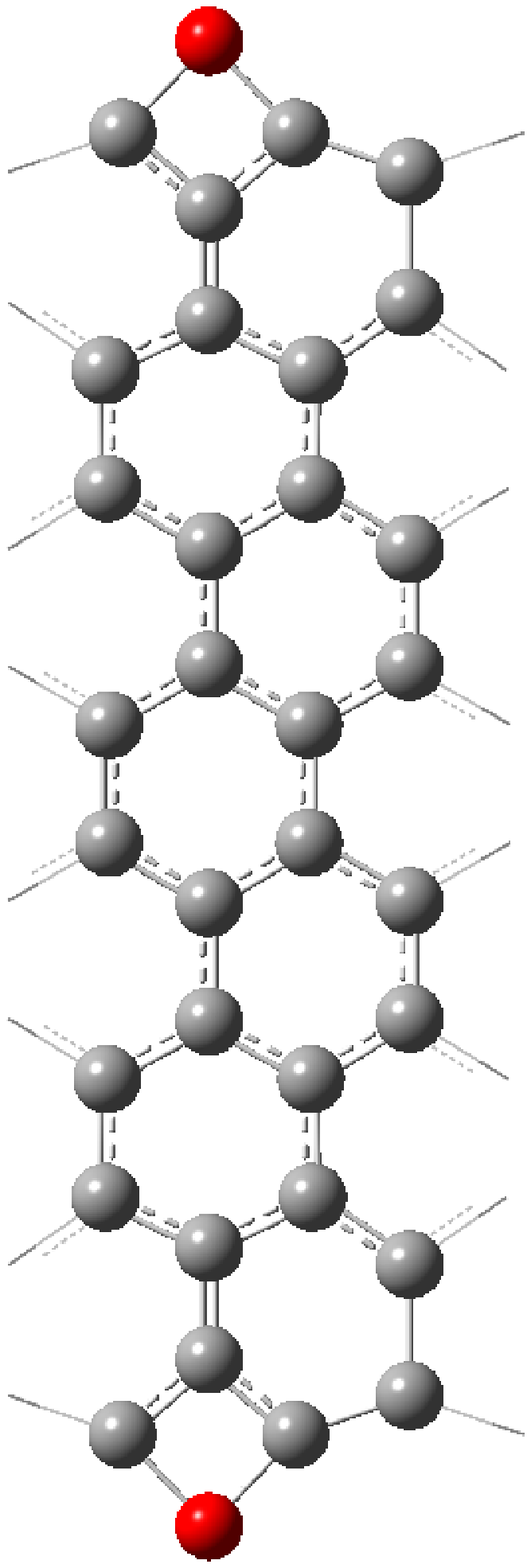}      &
\epsfxsize=1.30cm \epsffile{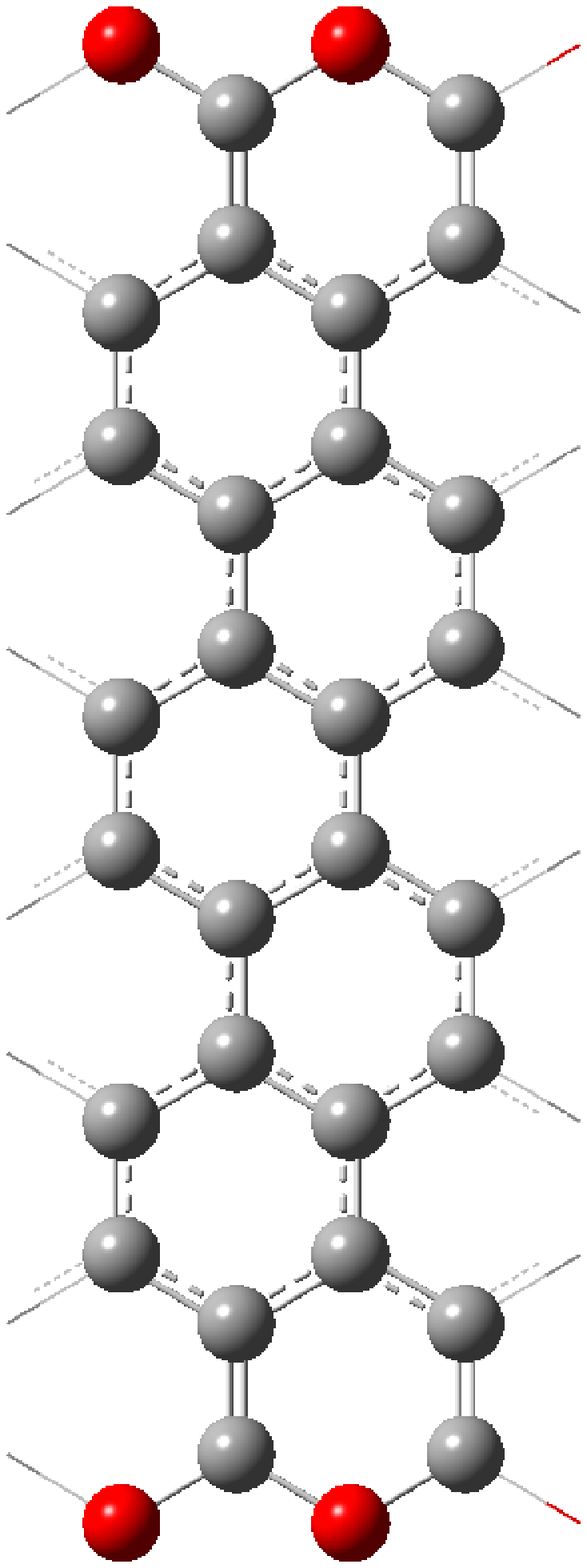} \\
Hydroxyl &
Hydroxyl &
Carboxyl &
Ketone   &
Ether    &
Ether    \\
\raisebox{0.5ex}[0pt]{(I)}  & 
\raisebox{0.5ex}[0pt]{(II)} & 
      &
      & 
\raisebox{0.5ex}[0pt]{(I)} & 
\raisebox{0.5ex}[0pt]{(II)}
\end{tabular} 
\end{center}
\caption{Optimized unit cell geometries of different edge oxidation schemes
studied in this work. Color code: red - oxygen atoms, gray - carbon
atoms, white - hydrogen atoms.}
\label{Fig: Oxidation Schemes}
\end{figure}
The importance of edge states was demonstrated for the case of
arm-chair CNTs which present a metallic and
non-magnetic~\cite{saito_book, dress-book} character in their tubular
form, but when unrolled into the corresponding zig-zag GNRs, they were
predicted to become semiconducting~\cite{Nakada1998, Son2006} with a
spin polarized~\cite{Fujita1996, Wakabayashi1998, Wakabayashi1999,
Kusakabe2003, Yamashiro2003, Lee2005, Son2006} ground state.  This
ground state is characterized by opposite spin orientations of
localized electronic states at the two edges of the GNR, which couple
through the graphene backbone via an antiferromagnetic (AF)
arrangement of spins on adjacent atomic sites.

Recently, Son {\em et al.}~\cite{Son2006} have shown that upon the
application of an electric field, an opposite local gating effect of
the spin states on the two edges of the ribbon may occur.  The
in-plane field (perpendicular to the main axis of the ribbon) drives
the system into a half-metallic state where one spin flavor exhibits a
metallic behavior, while the opposite spins experience an increase in
the energy gap.  Apart from the interesting interplay between the
electric field and the spin degree of freedom, this phenomenon is
important from a technological standpoint since such a system could
serve as a perfect spin filter in future nanospintronic devices.

Most of the theoretical studies on GNRs,\cite{Fujita1996, Nakada1996,
Wakabayashi1998, Nakada1998, Wakabayashi1999, Miyamoto1999, Kawai2000,
Kusakabe2003, Yamashiro2003, Lee2005, Peres2006, Ezawa2006,
Barone2006, Wang2007, White2007, Brey2007, Gunlycke2007, Pisani2007}
including those on the half-metallic systems,\cite{Son2006,Son2006-2}
address either bare or hydrogen terminated ribbons.  Nevertheless,
during standard GNRs fabrication processes, it is commonly assumed
that the ribbons edges become oxidized.~\cite{He1998, Lerf1998,
Radovic2005, Stankovich2006} Since many of the unique properties of
GNRs are associated with edge states, edge chemistry in general and
oxidation in particular, can significantly alter the electronic
properties of the ribbons.  It is the purpose of this letter to study,
for the first time, the influence of edge oxidation on the relative
stability, the electronic properties, and the half-metallic nature of
zigzag graphene nanoribbons, from {\em first-principles}.

To this end, we study six different oxidation schemes of a $1.8$~nm
wide zigzag GNRs including hydroxylation, carboxylation, ketonation,
and etheration, as shown in Figure~\ref{Fig: Oxidation Schemes}.  All
the calculations presented in this work were carried out using
periodic boundary conditions as implemented in the development version
of the {\it Gaussian} suite of programs.~\cite{gdv_short,pbca} Fully
relaxed geometries were obtained using the PBE realization of the
generalized gradient approximation~\cite{pbe_1,pbe_2} and the
polarized 6-31G** Gaussian basis set~\cite{Hariharan1973} for the
non-magnetic (closed-shell) state.  The electronic structures were
then re-evaluated using the screened exchange hybrid density
functional, HSE,~\cite{hse, hse-errata} which has been tested in a
wide variety of materials and has been shown to accurately reproduce
experimental band gaps~\cite{hse-bulk,hse-bulk2} and first and second
optical excitation energies in metallic and semiconducting
SWNTs.~\cite{chirals,metallic} Furthermore, the inclusion of
short-range exact-exchange in the HSE functional makes it suitable to
treat electronic localization effects~\cite{Kudin2002, Prodan2005,
Prodan2006, Hay2006, Kasinathan2006} which are known to be important
in this type of materials.~\cite{Kobayashi1993, Fujita1996,
Nakada1996, Wakabayashi1998, Nakada1998, Wakabayashi1999,
Miyamoto1999, Kawai2000, Okada2001, Kusakabe2003, Yamashiro2003,
Niimi2005, Kobayashi2005, Lee2005, Son2006, Son2006-2, Niimi2006,
Kobayashi2006}

We start by studying the relative stability of the different oxidized
ribbons.  As these structures have different chemical compositions,
the cohesive energy per atom does not provide a suitable measure for
the comparison of their relative stability.  Therefore, we adopt the
approach customary used in binary phase thermodynamics to account for
chemical composition and utilized previously to qualitatively analyze
the relative stability of endohedral silicon
nanowires~\cite{Dumitrica2004} and arm-chair GNRs.~\cite{Barone2006}
Within this approach one defines a Gibbs free energy of formation
$\delta G$ for a GNR as:
\begin{equation}
\delta G(\chi) = E(\chi) - \chi_{\sml H} \mu_{\sml H} - \chi_{\sml O} \mu_{\sml O} - \chi_{\sml C} \mu_{\sml C},
\label{Eq: Free energy}
\end{equation}
where $E(\chi)$ is the cohesive energy per atom of the GNR, $\chi_i$
is the molar fraction of atom $i$ ($i$=C,O,H) in the ribbon,
satisfying the relation $\Sigma_i\chi_i=1$, and $\mu_i$ is the
chemical potential of the constituent $i$ at a given state.  We choose
$\mu_{\sml H}$ as the binding cohesive energy per atom of the singlet
ground state of the H$_2$ molecule, $\mu_{\sml O}$ as the cohesive
energy per atom of the triplet ground state of the O$_2$ molecule, and
$\mu_{\sml C}$ as the cohesive energy per atom of a single graphene
sheet.  This definition allows for a direct energy comparison between
oxidized nanoribbons with different compositions, where negative
values represent stable structures with respect to the constituents.
It should be stressed that this treatment gives a qualitative
assessment of the relative stability while neglecting thermal and
substrate effects and zero point energy corrections.

\begin{figure}
\centerline{\includegraphics[width=3.5in]{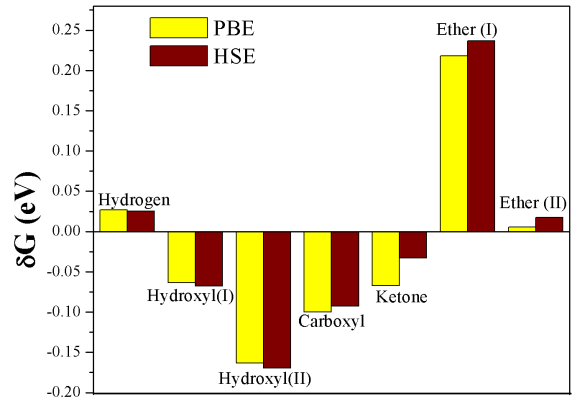}}
\caption{Ground state relative stabilities of the different oxidation schemes
studied (see Fig.~\ref{Fig: Oxidation Schemes}) obtained via
Eq.~\ref{Eq: Free energy} at the PBE and HSE/6-31G** levels of theory.
Negative values indicate stable structures with respect to the
constituents.}
\label{Fig: Stability}
\end{figure}

In Figure~\ref{Fig: Stability} we present the relative stability of
the ground states of the different oxidation schemes studied.  The
fully hydrogenated ribbons, as well as the two etherated ribbon
configurations, are found to be less stable than their corresponding
constituents.  This is consistent with previous calculations on bare
and hydrogen terminated GNRs.~\cite{Barone2006} Nevertheless, the
chemical passivation via hydroxylation, ketonization and carboxylation
leads to considerable energetic stabilization of the structure of the
ribbons.  We find that the most stable structure corresponds to the
fully hydroxylated ribbon.  This enhanced stability is attributed to
the hydrogen bonds formed between adjacent hydroxyl groups.

\begin{figure}
\begin{center}
\begin{tabular}{ccccc}
 \epsfxsize=1.50cm  \epsffile{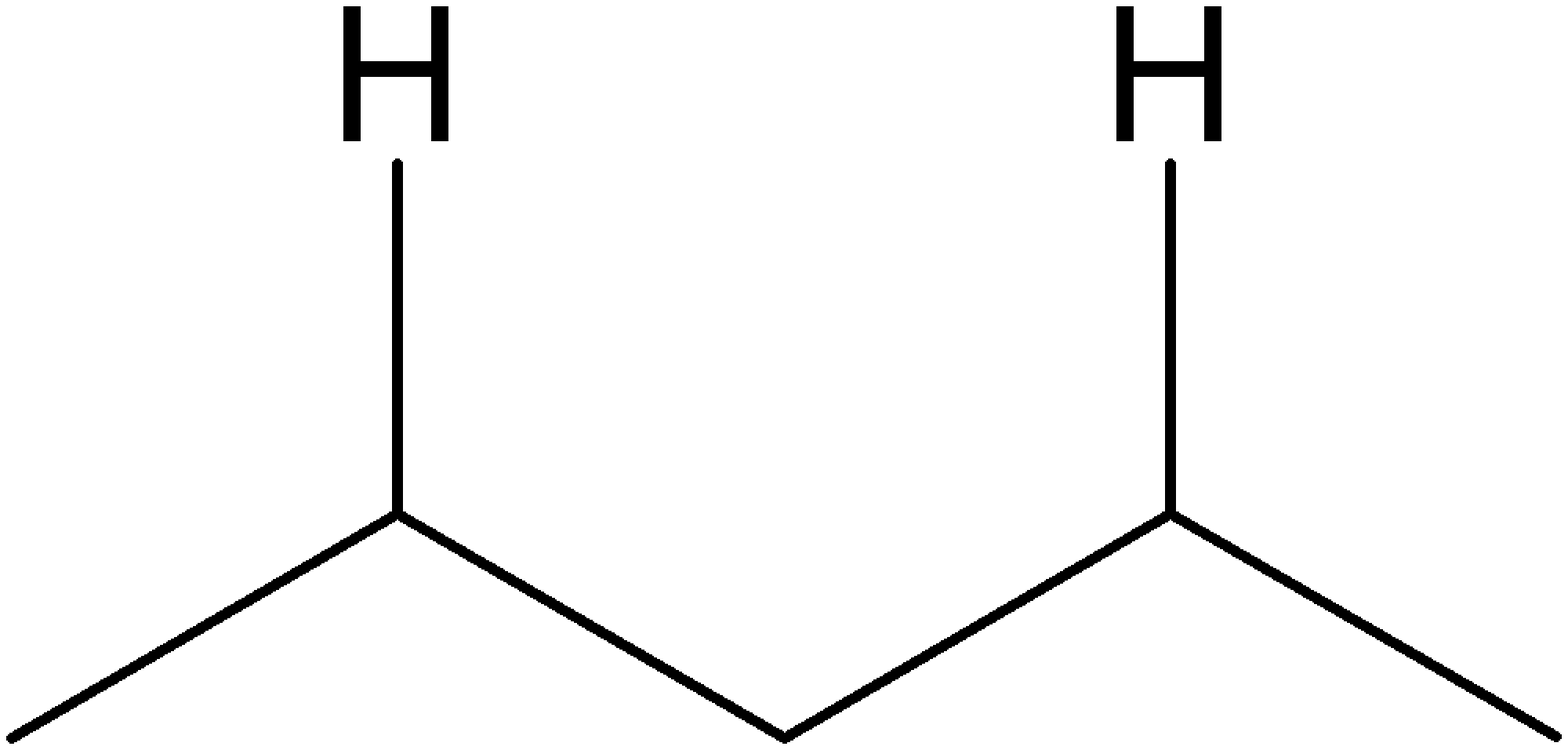} &
 \epsfxsize=1.50cm  \epsffile{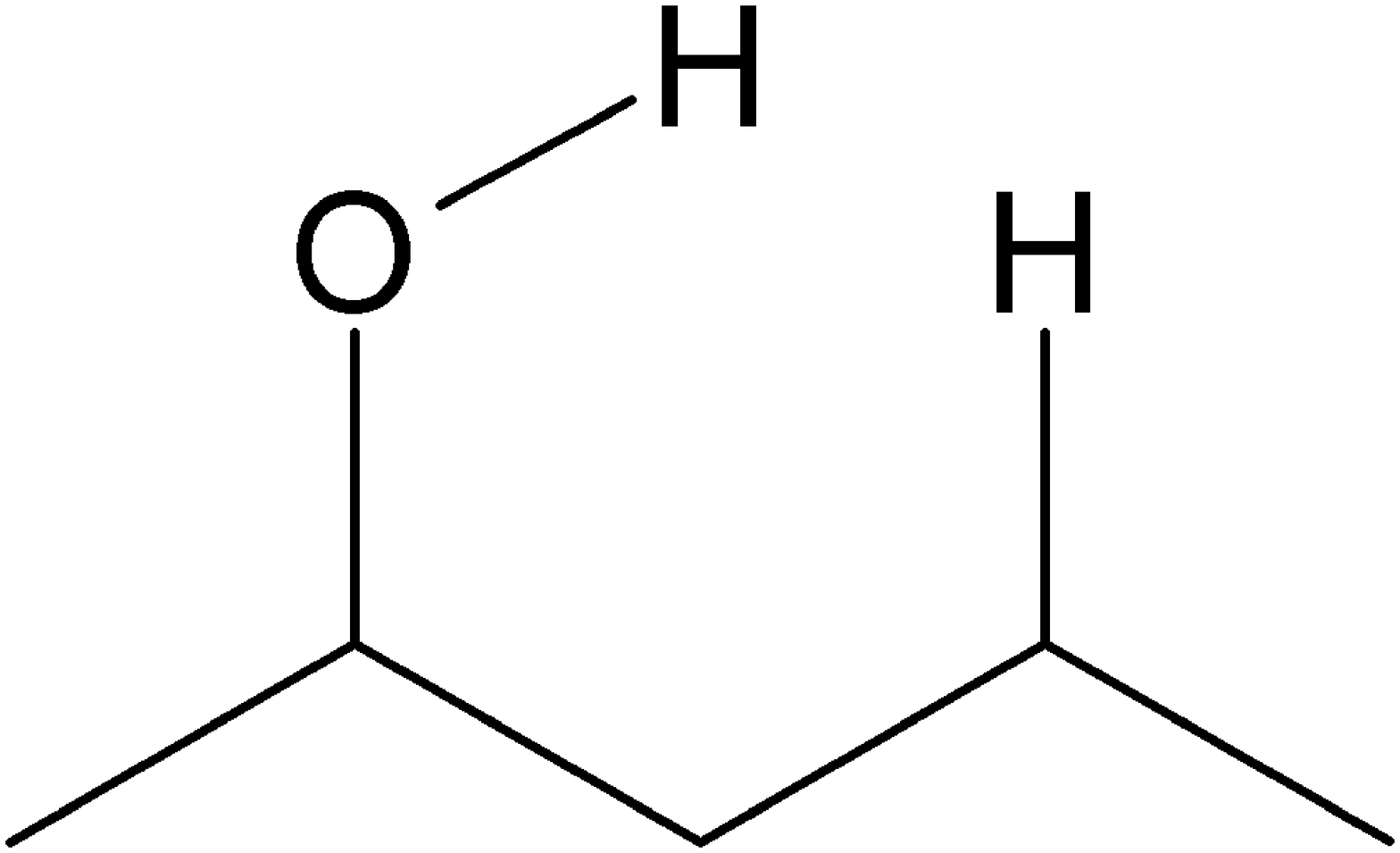}  &
 \epsfxsize=1.50cm  \epsffile{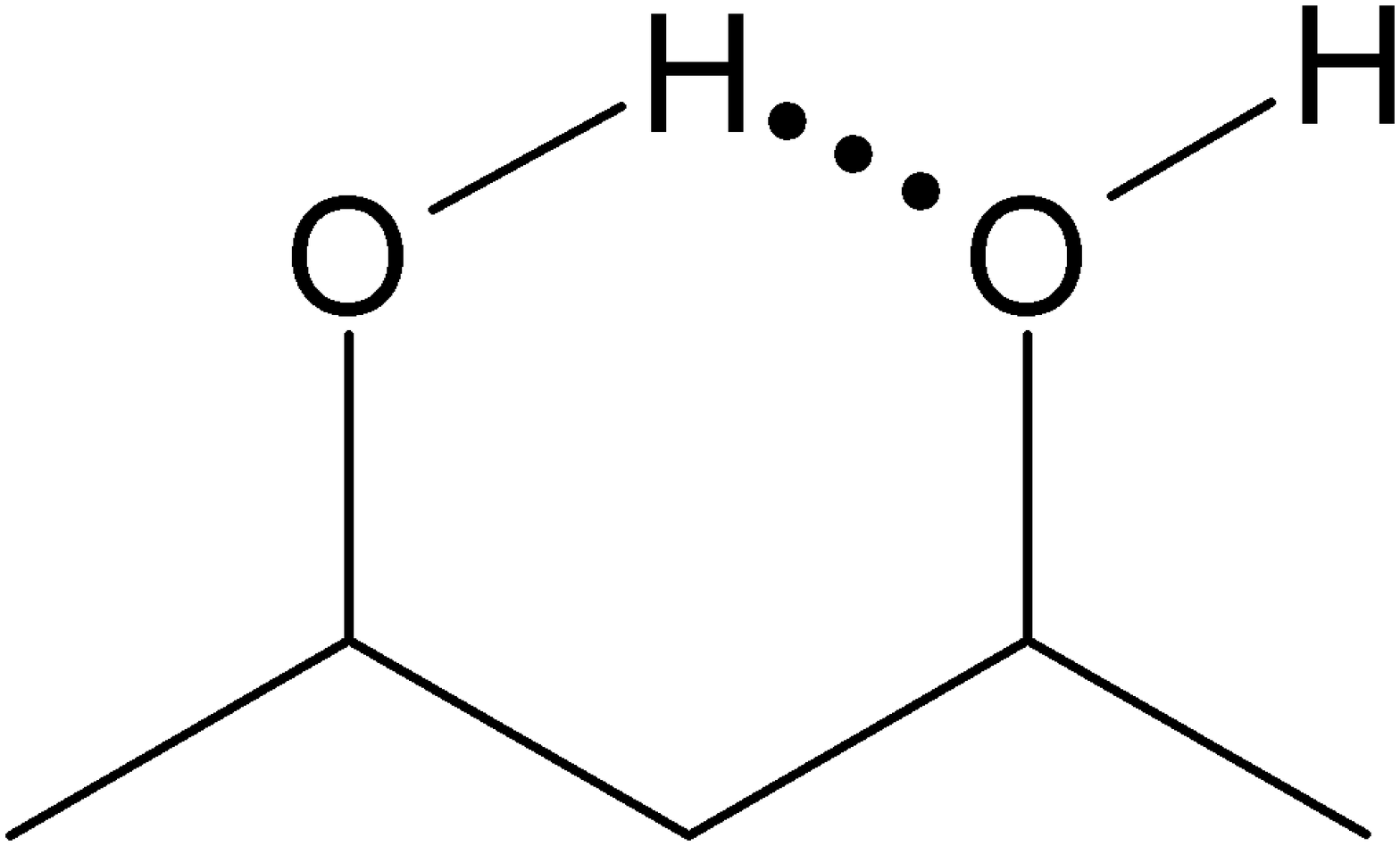}  &
 \epsfxsize=1.50cm  \epsffile{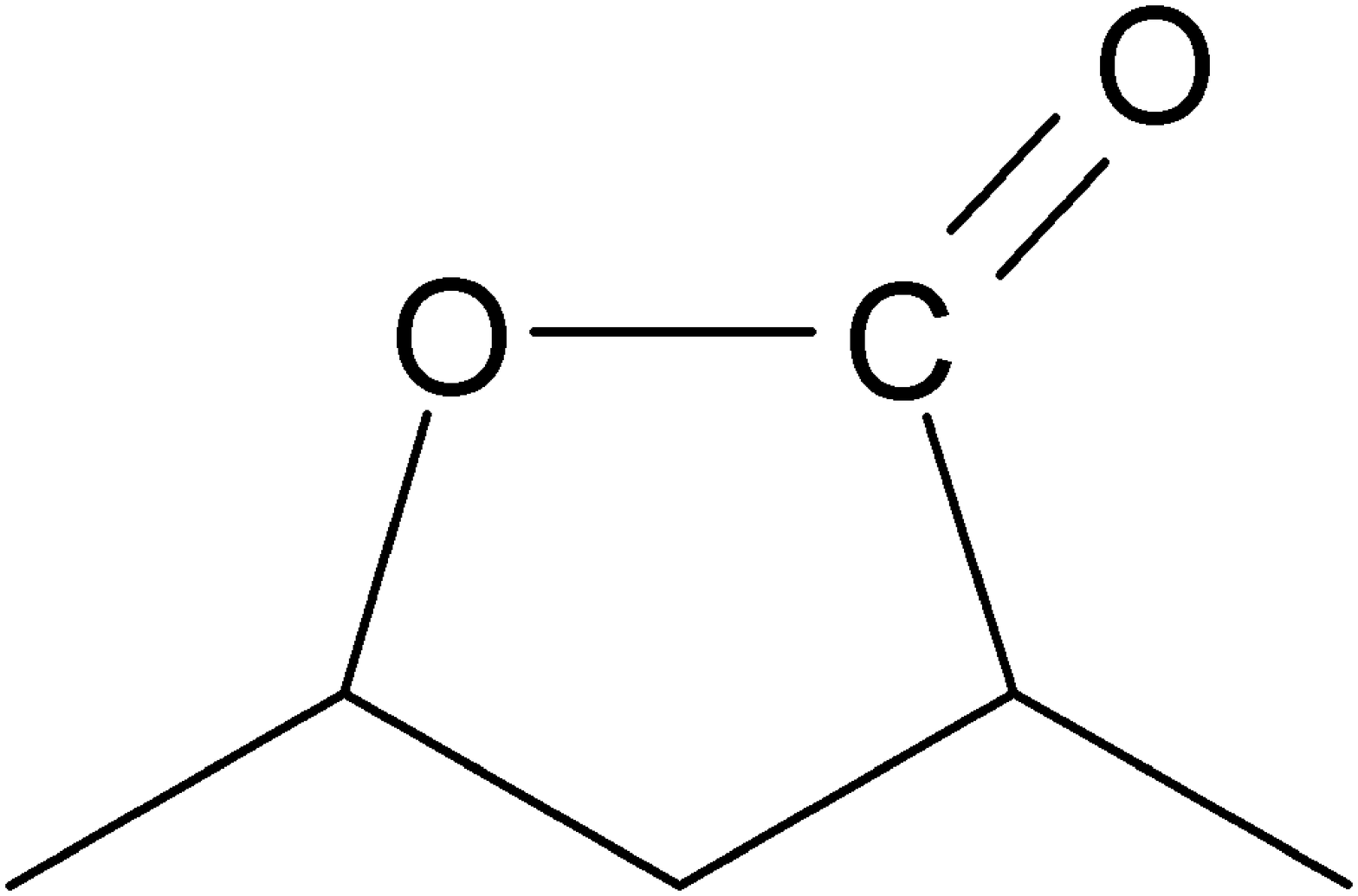}  &
 \epsfxsize=1.50cm  \epsffile{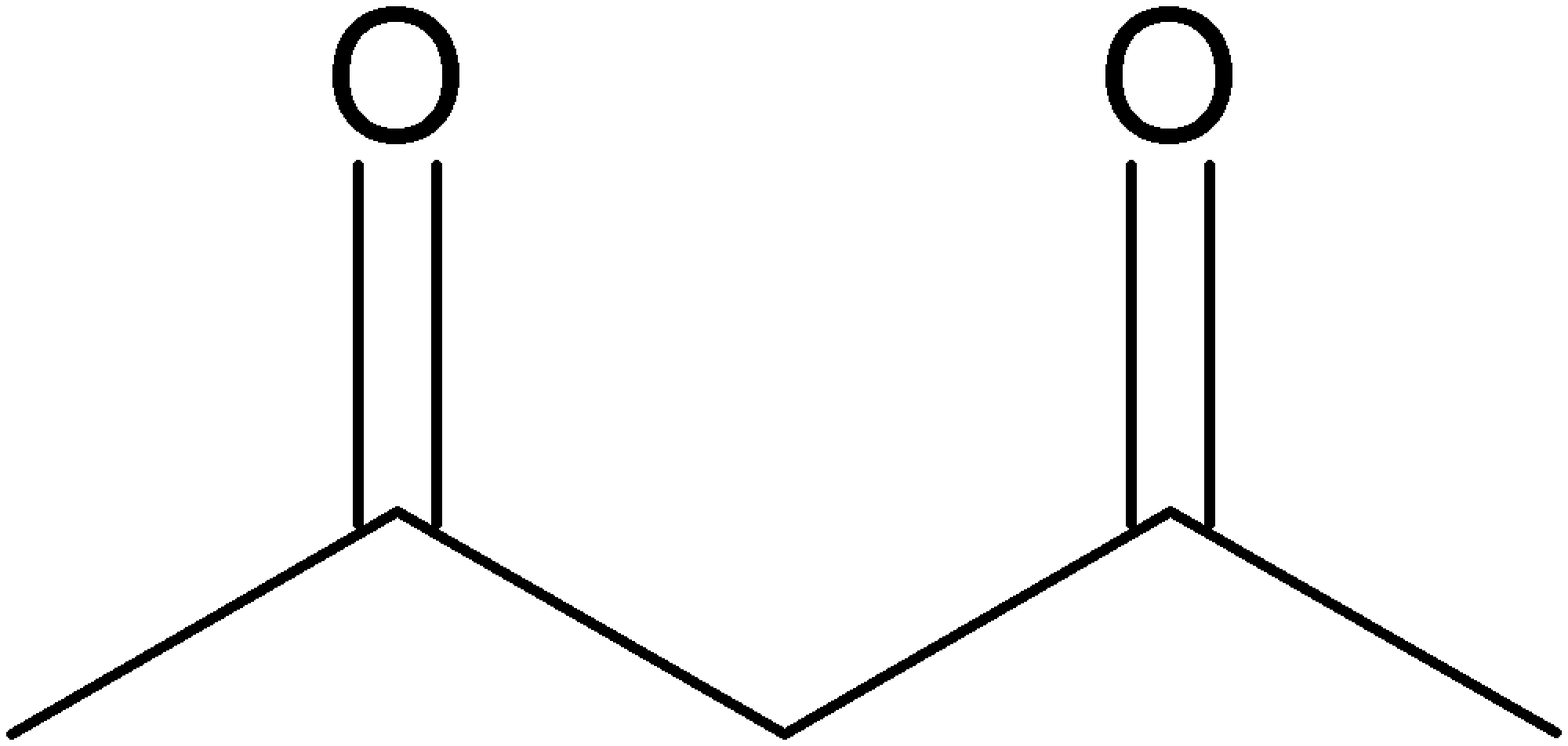}  \\
\multicolumn{5}{c}{\epsfxsize=8.00cm \epsffile{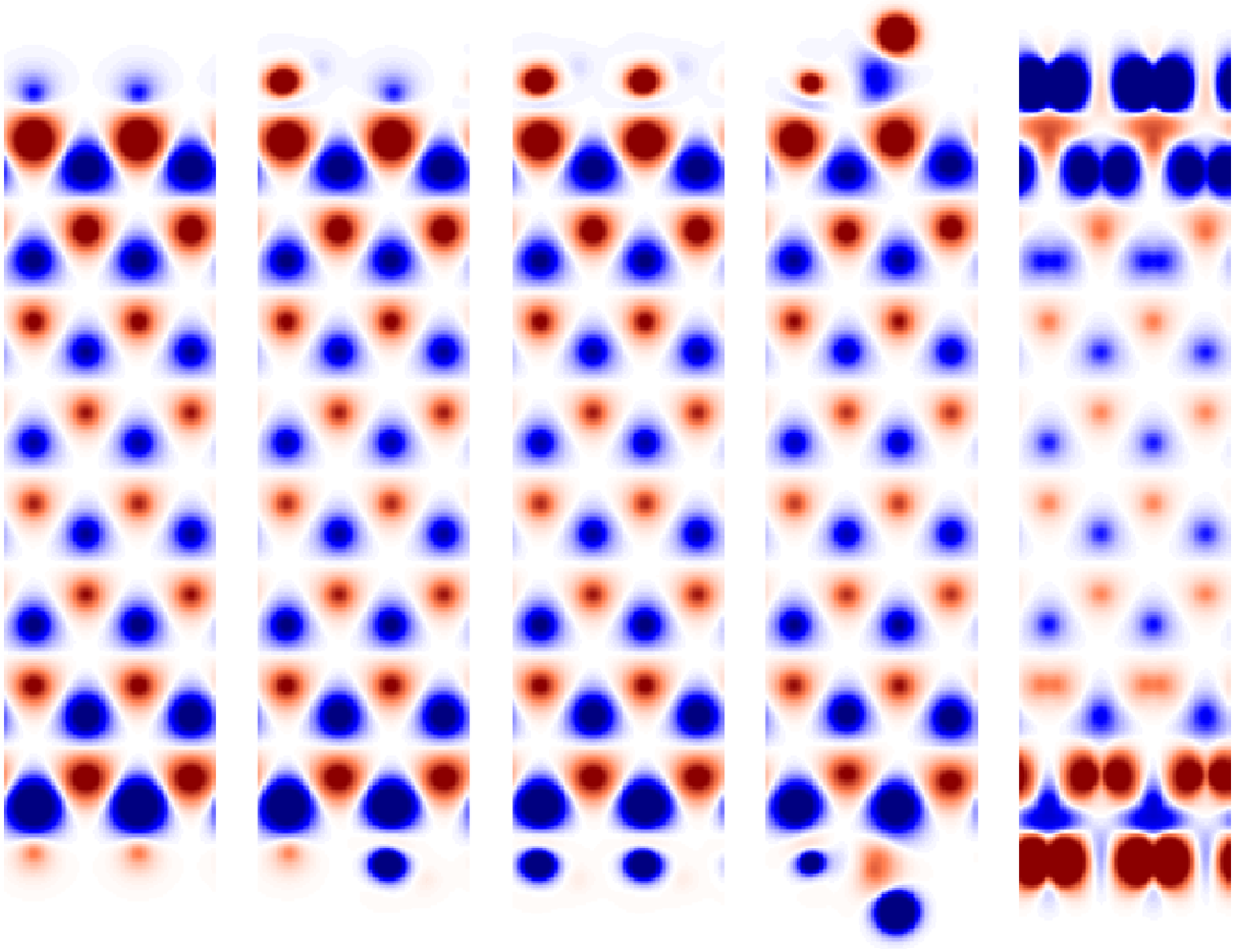}}  \\
\multicolumn{5}{l}{{\Huge $\longrightarrow$}} \\
\multicolumn{5}{l}{Periodic Direction}
\end{tabular}
\end{center}
\caption{Map of the HSE ground state spin densities $0.16$~{\AA}
  above the surface of the ribbon for the four most stable
  oxidation schemes studied.  The fully hydrogenated system is also
  included for comparison.  Color code: Red $\alpha$ spin density;
  blue $\beta$ spin density.  A scheme of the different oxidation
  groups is presented for clarity.\label{Fig: Spin Density} }
\end{figure}

We now turn to study the electronic properties of the four most stable
structures.  In Figure~\ref{Fig: Spin Density} we show the ground
state spin density of the fully hydrogenated, ketonated, hydroxylated
and carboxylated structures, obtained with the HSE
functional.~\cite{PBE-Footnote} Similar to the fully hydrogenated
GNR~\cite{Son2006}, all the depicted oxidized structures exhibit a
spin polarized ground state where the spin magnetization on the
opposite edges of the ribbons are aligned anti-parallel.  For the
hydroxylated and the carboxylated ribbons, the presence of the
oxidation groups slightly changes the absolute value of the spin
polarization at the edge carbon atoms. However, the overall
magnetization density resembles that of the fully hydrogenated GNR. A
different picture arises for the ketonated system, where the oxygen
$p$-electrons participate in the $\pi$ system of the ribbon and
therefore present considerable spin polarization, which results in a
qualitatively different spin density map compared to that of the fully
hydrogenated system.

\input{epsf}

In Table~\ref{Tab: Bandgaps} we present the bandgaps and the energy
differences per unit cell between the AF ground state and the
above-lying ferromagnetic (FM) state (where the edges of the ribbon
bare parallel spin polarization) for the hydrogenated, and stable
oxidized structures.  All studied oxidized structures, except for the
ketonated ribbons, have a bandgap comparable to that of the fully
hydrogenated ribbon, indicating that their electronic character is
only slightly changed upon oxidation.  Furthermore, the energy
difference between the AF ground state and the FM state is above room
temperature and similar to that of the hydrogenated ribbon.  The
ketonated ribbon, on the other hand, exhibits a vanishing bandgap and
a very small energy difference between the AF ground state and the FM
state, suggesting that it will be extremely difficult to observe the
half-metallic nature of the ketonated ribbons even at low
temperatures.
\begin{table}
 \caption{HSE/6-31G** AF ground state bandgaps (eV) and energy
differences between the FM and AF states (meV/unit cell) of the
studied GNRs. \label{Tab: Bandgaps}}
\begin{ruledtabular}
\begin{tabular}{lcc}
                     & $E_g$ & $E_{\sml FM}-E_{\sml AF}$ \\ \hline
   Hydrogenated      &   1.05     & 42 \\ 
   Hydroxylated (I)  &   0.99     & 39 \\ 
   Hydroxylated (II) &   0.90     & 38 \\ 
   Carboxylated      &   0.77     & 33 \\ 
   Ketonated         &   0.03     & 5 \\ 
\end{tabular}
\end{ruledtabular}
\end{table}

\begin{figure}
\centerline{\includegraphics[width=3.20in]{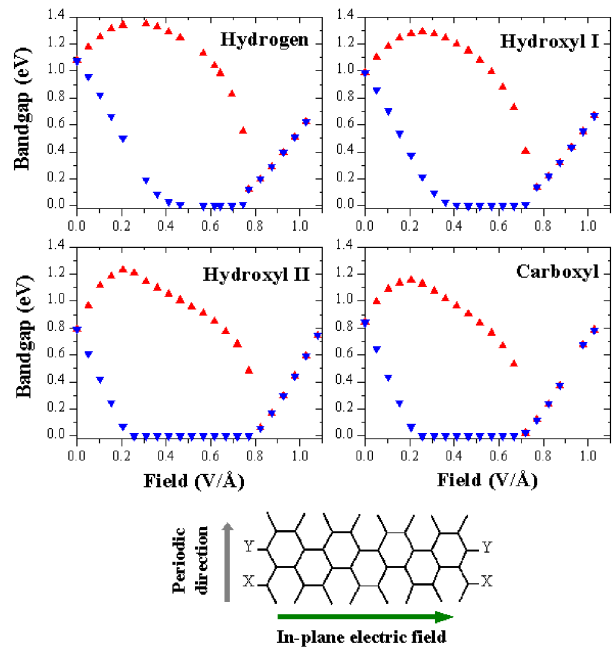}}
\caption{Electric field effect on the spin polarized bandgap of oxidized
GNRs. Red and blue triangles stand for the two different spin
polarization bandgaps.  The values are calculated using the HSE
functional and the 3-21G basis set.~\cite{Binkley1980} A scheme of the
direction of the applied electric field is presented for clarity,
where X and Y represent the oxidizing groups.}
\label{Fig: Electric field}
\end{figure}

The analysis presented above suggests that, with the exception of the
ketonated systems, all oxidation schemes have little effect on the
electronic character of the GNRs and therefore, one would expect these
systems to behave as half-metals under the influence of an external
electric field.  To verify this assumption we have calculated the
bandgap of the $\alpha$ and $\beta$ spin channels as a function of the
intensity of a transverse in-plane, uniform, and static electric
field.  In Figure~\ref{Fig: Electric field} we plot the spin resolved
bandgaps as a function of the field intensity for the fully
hydrogenated, hydroxylated, and carboxylated ribbons.
In the absence of an electric field, the $\alpha$ and $\beta$ bandgaps
are degenerate for all the studied systems.  Upon the application of a
low intensity electric field, a splitting occurs where the bandgap of
one spin flavor increases and that of the opposite spin flavor
decreases.~\cite{Son2006} The bandgap splitting increases
monotonically with the field intensity up to a point where the system
becomes half-metallic.  Further increase in the external electric
field intensity results in a decrease of the bandgap splitting up to a
point where the systems become non-magnetic.

It is interesting to note that when the edges of the ribbon are fully
or partially hydrogenated, the field intensity needed to switch the
system to the half-metallic regime is $0.4$~V/\AA ~and the range at
which the half-metallic behavior is maintained is $0.3$~V/\AA.
Nevertheless, when the edges are fully oxidized, the system turns
half-metallic at a lower field intensity ($0.2$~V/\AA) and the range
of half-metallic behavior is doubled to $0.6$~V/\AA. As shown
previously in Ref.~\onlinecite{Son2006}, the intensity of the field
needed to achieve half-metallicity decreases with the ribbon width.
For a given width, the decrease of the field intensity needed to
obtain the half-metallic behavior combined with the enhancement of the
range at which this behavior is maintained, implies that edge
oxidation is critical for the fabrication of robust and chemically
stable spin filter devices out of zigzag GNRs.

\begin{figure}
\centerline{\includegraphics[width=2.75in, angle=-90]{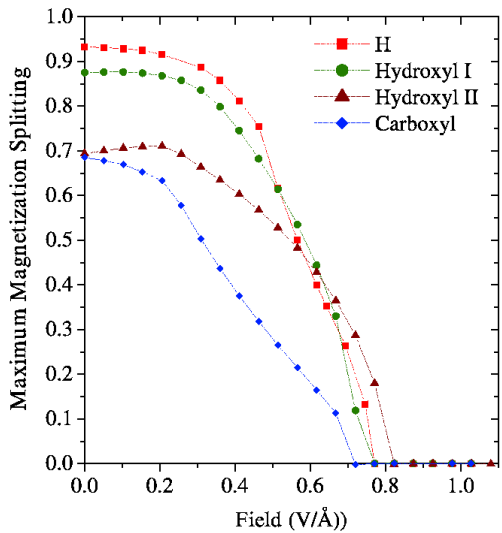}}
\caption{Largest magnetization splitting between the two edges of the ribbon as
a function of the electric field intensity.  Calculations performed on
the HSE/3-21G level of theory.}
\label{Fig: Magnetism}
\end{figure}

It is relevant to investigate how the spin magnetization depends on the intensity
of the applied electric field.
To this end, in Figure~\ref{Fig: Magnetism} we present the maximum
magnetization splitting as a function of the electric field, i.e. the difference between 
the maximum $\alpha$ and $\beta$ Mulliken atomic spin densities.

For the four systems studied, at the low electric field regime, the
field has minor influence on the magnetization splitting.  Once the
systems become half-metallic, the magnetization drops sharply with the
applied field up to a point where it fully quenches.  An explanation
for this behavior can be found by realizing that once the systems
become half-metallic, the electric polarizability of the metallic spin
channel increases considerably, allowing for a more effective spin
compensation between the opposite edges of the ribbon. This results in
an overall quenching of the spin magnetization. The electric field
above which all the systems become diamagnetic is $0.7$-$0.8$~V/{\AA}
for ribbons of about 1.8~nm wide.

In summary, we have presented a detailed study of the electronic
properties and relative stabilities of several edge-oxidized zigzag
GNRs.  The structure of the oxidized ribbons is found to be
stabilized with respect to the fully-hydrogenated counterparts except
for the case of the etheric groups.  Notably, all the stable oxidized
structures studied maintain a spin polarized ground state with
antiferromagnetic ordering.  The above-lying ferromagnetic
arrangement is expected to become accessible only beyond room
temperature.  Apart from the ketonated ribbons, all oxidized systems
present a lower onset field and a wider electric field range for which
half-metallic behavior is predicted.  This suggest that edge oxidation
is important for the design of efficient and robust spintronic devices
based on GNRs.  We have also shown that once the half-metallic state
is reached, further increase of the external electric field produces a
rapid decrease in the spin magnetization up to a point where all the
systems become non-magnetic.


{\Large \bf  Acknowledgments}

This work was supported by NSF Award Number CHE-0457030 and the Welch
Foundation. Calculations were performed in part on the Rice Terascale
Cluster funded by NSF under Grant EIA-0216467, Intel, and HP.
O.H. would like to thank the generous financial support of the
Rothschild and Fulbright foundations.


\bibliographystyle{./jcp2.bst} \bibliography{OXY-ZZ-GNR}

\begin{thebibliography}{10}
\bibitem{PBE-Footnote}
For some of the oxidation schemes, when using the PBE functional, we could not
  identify an anti-ferromagnetic arrangement and a non-magnetic ground state
  was found.
\end{thebibliography}
\end{document}